\documentclass[cameraready]{Interspeech}
\usepackage{microtype}
\usepackage{graphicx}
\usepackage{subcaption}
\usepackage{booktabs}
\usepackage{hyperref}
\usepackage{amsmath}
\usepackage{amssymb}
\usepackage{mathtools}
\usepackage{amsthm}
\usepackage{booktabs}
\usepackage{multirow}
\usepackage{xcolor}
\usepackage{array}
\usepackage{colortbl}
\usepackage{adjustbox}
\usepackage{caption}

\title{MDM-ASR: Bridging Accuracy and Efficiency in ASR with Diffusion-Based Non-Autoregressive Decoding}
\author[affiliation={1}, equalcontribution]{Hao}{Yen}
\author[affiliation={1}, equalcontribution]{Pin-Jui}{Ku}
\author[affiliation={3}]{Ante}{Juki\'{c}}
\author[affiliation={1,2}]{Sabato Marco}{Siniscalchi}

\address{
    $^1$ Georgia Institute of Technology, USA \\
    $^2$ Università degli Studi di Palermo, Italy \\
    $^3$ NVIDIA, USA
}

\email{rick.yen@gatech.edu, pku9@gatech.edu, ajukic@nvidia.com}


\usepackage{comment}

\begin{document}

\maketitle

\begin{abstract}
In sequence-to-sequence Transformer ASR, autoregressive (AR) models achieve strong accuracy but suffer from slow decoding, while non-autoregressive (NAR) models enable parallel decoding at the cost of degraded performance. We propose a principled NAR ASR framework based on Masked Diffusion Models to significantly reduce this gap. A pre-trained speech encoder is coupled with a Transformer diffusion decoder conditioned on acoustic features and partially masked transcripts for parallel token prediction. To mitigate the training–inference mismatch, we introduce {\em Iterative Self-Correction Training} that exposes the model to its own intermediate predictions. We also design an {\em Position-Biased Entropy-Bounded Confidence-based} sampler with positional bias to further boost results. Experiments across multiple benchmarks demonstrate consistent gains over prior NAR models and competitive performance with strong AR baselines, while retaining parallel decoding efficiency.

\end{abstract}
\section{Introduction}
\label{sec:intro}

Automatic Speech Recognition (ASR)~\cite{Li2022} has become a cornerstone technology, powering applications from smart home devices and virtual assistants to real-time transcription. Over the past decade, deep learning advances have greatly improved ASR accuracy and robustness. Connectionist Temporal Classification (CTC)~\cite{Graves2006,Graves2014}, which enables alignment-free training by mapping input frames directly to output sequences, initially revolutionized ASR. CTC-based models are non-autoregressive (NAR), allowing parallel token decoding, and typically combine strong encoders, such as Wav2Vec2.0~\cite{Baevski2020}, HuBERT~\cite{Hsu2021}, or MMS~\cite{Pratap2023}, with a CTC prediction head and loss. This design enables very fast decoding~\cite{Peng2024,Peng2025} across benchmarks~\cite{Srivastav2025}. However, the strong conditional independence assumption between output tokens limits the modeling of rich linguistic dependencies, often resulting in performance degradation~\cite{Watanabe2017} and preventing CTC from dominating state-of-the-art ASR. 


To address CTC’s limitations, Conformer-based transducers~\cite{Huang2020,Tripathi2020,Chen2020,Zhang2020} combine a Conformer~\cite{Gulati2020} encoder with an autoregressive (AR) token predictor via a joint network, making them well suited for streaming ASR. Another class of AR models are sequence-to-sequence (seq2seq) Transformer ASR systems~\cite{Vaswani2017,Dong2018}, which leverage an encoder-decoder architecture~\cite{Chorowski2015,Chan2016} with cross-attention to implicitly align speech frames and output tokens. This approach has produced highly competitive fully Transformer-based systems, including Whisper~\cite{Radford2023} and Canary~\cite{Zelasko2025}, achieving strong accuracy across multiple languages and domains.
Despite their effectiveness, AR models generate tokens sequentially, so inference time grows linearly with output length. Maintaining long-range context during decoding further increases computational cost, limiting their suitability for real-time or large-scale applications~\cite{Li2022,Watanabe2017}.

In text generation, NAR models have recently gained considerable attention, demonstrating their potential to be a valid alternative to AR solutions.
In particular, models based on diffusion~\cite{Sohl2015,Ho2020} and flow-matching~\cite{Gat2024,Shaul2025} frameworks enable parallel sequence generation and allow a flexible trade-off between accuracy and efficiency through iterative refinement. Among them, masked diffusion models (MDMs)~\cite{Austin2021,Hoogeboom2021,Campbell2022,He2023} have emerged as a promising solution for sequence modeling.
Instead of generating tokens one by one, MDMs start from a masked sequence and iteratively predict multiple tokens in parallel.
This bidirectional refinement process allows the model to use both past and future contexts at every step, leading to improved modeling capacity while significantly reducing inference latency. MDMs have shown promise in unconditional text generation~\cite{Zheng2024,Sahoo2024,Ye2025} and zero-shot evaluation~\cite{Nie2025,Sahoo2025} for language modeling tasks. 
Given that ASR fundamentally performs text generation conditioned on speech, these advances suggest that diffusion-based NAR generation may also be beneficial for speech recognition. 
However, the scalability of MDMs and their effectiveness for ASR have not been fully explored. Moreover, very recent studies on diffusion-based ASR~\cite{Baas2022,Yeh2024,Kwon2025} and flow-matching ASR~\cite{Navon2025} report a substantial performance gap compared with strong AR ASR models.
    
In this work, we study NAR modeling within the standard seq2seq Transformer ASR framework and seek to close the performance gap between AR and NAR approaches. To this end, we introduce an audio-conditioned masked diffusion decoder that replaces conventional left-to-right AR decoding while preserving the encoder–decoder architecture.
Our model consists of a strong pre-trained speech encoder to extract high-level acoustic representations, which are then consumed by a Transformer-based discrete diffusion decoder that iteratively refines masked token sequences. 
At each diffusion decoding step, the decoder jointly conditions on the complete acoustic embedding and the partially masked text sequence, enabling parallel token updates and bidirectional context modeling across the entire output. 
Unlike prior diffusion ASR attempts that couple separately trained speech and text components via an additional fusion module~\cite{Kwon2025,Yeh2024}, or recent flow-matching ASR efforts that introduce an additional auxiliary modeling~\cite{Navon2025}, our model builds directly on a standard encoder–decoder ASR architecture and replaces sequential decoding with a masked diffusion generation process. 
Moreover, unlike earlier BERT-style masked refinement ASR~\cite{Chen2021}, which relies on ad-hoc objectives without an explicit generative interpretation, we employ a discrete diffusion formulation with a training objective derived from the forward corruption process. The resulting diffusion loss, including the noise-dependent reweighting term, provides each refinement step with a clear probabilistic interpretation. In addition, it defines a theoretically grounded multi-step generative model, whose inference procedure is fully consistent with the training process. 


Building on this framework, we propose several practical techniques to further improve performance. First, {\em Iterative Self-Correction Training} (ISCT) reduces the mismatch between training and inference by exposing the model to intermediate decoding states. Second, we explore multiple inference strategies, including the {\em Entropy-Bounded Confidence} (EB-Conf) sampler for stable and efficient decoding, and the {\em Position-Biased EB-Conf} (PBEB-Conf) sampler to incorporate positional bias. Experiments on four benchmark English datasets show that our model achieves competitive accuracy with strong AR systems while delivering significantly faster decoding and consistently outperforming existing generative NAR ASR models. On multilingual tasks, it demonstrates robust performance across languages. Comprehensive ablation studies further examine model scaling, ISCT, and inference strategies, providing clear insights into system design and empirical behavior.

\section{Related Work}
\label{sec:RelatedWork}

\subsection{Masked Diffusion Model}
\label{sec:masked_diffusion}

Discrete diffusion models \cite{Austin2021,Hoogeboom2021,Nichol2021,Lou2024,Liu2025ddpd} extend diffusion-based generative modeling to categorical variables, enabling diffusion-style learning in discrete token spaces such as text. In contrast to continuous diffusion models that progressively inject Gaussian noise into signals, discrete diffusion defines a stochastic corruption process over a finite vocabulary and learns a corresponding reverse-time denoising process.

Let $\boldsymbol{x}$ be a one-hot vector in $\{0, 1\}^{|\mathcal{V}|}$  representing a clean token from the vocabulary $\mathcal{V}$ with size $|\mathcal{V}|$. A forward noising process constructs a sequence of latent variables $\boldsymbol{z}_t$, indexed by a continuous diffusion time
$t \in [0,1]$. A commonly used formulation is the \emph{masked diffusion process}, in which tokens are progressively replaced by an absorbing mask state, denoted with a special \texttt{[MASK]} token in $\mathcal{V}$. The marginal forward distribution at time $t$ is defined as
\begin{equation}
q(\boldsymbol{z}_t \mid \boldsymbol{x})=
\mathrm{Cat}\left(
\boldsymbol{z}_t;
\alpha_t \boldsymbol{x} + (1-\alpha_t) \boldsymbol{m}\right),
\label{eq:forward_process}
\end{equation}
where $\mathrm{Cat}(\cdot;\boldsymbol{p})$ denotes a categorical distribution over $|\mathcal{V}|$ classes with probabilities $\boldsymbol{p} \in \Delta^{|\mathcal{V}|}$, where $\Delta^{|\mathcal{V}|}$ denotes the $|\mathcal{V}|$-dimensional probability simplex that sums to 1, $\boldsymbol{m}$ is the one-hot vector corresponding to the mask token, and $\alpha_t : [0,1] \to [0,1]$ is a monotonically decreasing noise schedule satisfying $\alpha_0 \approx 1$ and $\alpha_1 \approx 0$. As $t$ increases, probability mass gradually shifts from the original token to the absorbing mask state. Under this absorbing-state construction, the conditional distribution of an intermediate latent variable $\boldsymbol{z}_s$ given a later variable $\boldsymbol{z}_t$ and the clean token $\boldsymbol{x}$, for $0 < s < t < 1$, admits a closed-form expression
\begin{equation}
q(\boldsymbol{z}_s \mid \boldsymbol{z}_t, \boldsymbol{x}) = \mathrm{Cat}\left(\boldsymbol{z}_s;
\frac{\alpha_s - \alpha_t}{1-\alpha_t} \boldsymbol{x} + \frac{1-\alpha_s}{1-\alpha_t}\boldsymbol{z}_t
\right) .
\label{eq:posterior}
\end{equation}
This posterior provides a convenient parameterization for defining the reverse-time denoising process.

The generative model learns a parameterized reverse transition $p_\theta(\boldsymbol{z}_s \mid \boldsymbol{z}_t)$ by replacing the unknown clean token $\boldsymbol{x}$ in~\eqref{eq:posterior} with a neural network prediction $\boldsymbol{x}_\theta(\boldsymbol{z}_t, t)$, which outputs a categorical distribution over the vocabulary $\mathcal{V}$ conditioned on the noisy input $\boldsymbol{z}_t$ and diffusion time $t$. In the masked diffusion setting, the reverse transition simplifies to the following piecewise form
\begin{equation}
\begin{aligned}
&p_\theta(\boldsymbol{z}_s \mid \boldsymbol{z}_t)
= q(\boldsymbol{z}_s \mid \boldsymbol{z}_t, \boldsymbol{x}_\theta(\boldsymbol{z}_t,t)) \\
&=\begin{cases}
\mathrm{Cat}(\boldsymbol{z}_s; \boldsymbol{z}_t), & \boldsymbol{z}_t \neq \boldsymbol{m} \\ \mathrm{Cat}\left( \boldsymbol{z}_s; \frac{(1-\alpha_s)\boldsymbol{m} + (\alpha_s-\alpha_t)\,\boldsymbol{x}_\theta(\boldsymbol{z}_t,t)}{1-\alpha_t} \right), & \boldsymbol{z}_t = \boldsymbol{m} 
\end{cases} \nonumber
\label{eq:reverse_process}
\end{aligned}
\end{equation}
That is, positions that are already unmasked remain unchanged, while masked positions are progressively unmasked using the model's predictions.

For a sequence of length $L$, the clean sequence can be written as $\boldsymbol{x}^{(1:L)} = \{\boldsymbol{x}^{(1)}, \ldots, \boldsymbol{x}^{(L)}\}$, and the corresponding masked sequence at time $t$ is denoted $\boldsymbol{z}_t^{(1:L)}$. The corruption process is applied independently across token positions. Training is performed by randomly sampling diffusion time $t$, generating corrupted sequences $\boldsymbol{z}_t^{(1:L)}$ using the forward process, and maximizing the likelihood of the original clean tokens under the model's reverse predictions. The training objective for the whole sequence can be written as
\begin{equation}
\mathcal L = \mathbb E_q \int_{0}^{1} \frac{\alpha_t'}{1-\alpha_t} \sum_{\ell=1}^{L} \log \left\langle
\boldsymbol{x}_\theta^{(\ell)}(\boldsymbol{z}_t^{(1:L)}, t), \boldsymbol{x}^{(\ell)} \right\rangle \, \mathrm{d}t .
\label{eq:training_objective}
\end{equation}
where $\alpha_t' = \frac{\mathrm{d}\alpha_t}{\mathrm{d}t}$ and $\boldsymbol{x}_\theta^{(\ell)}(\cdot)$ denotes the predicted probability for token at position $\ell$. This objective encourages accurate reconstruction of masked tokens across all noise levels, yielding a fully non-autoregressive generative model over discrete sequences.

\subsection{Diffusion-/Flow-matching based NAR ASR}
\label{sec:nar_asr}
Research on generative NAR models for ASR remains relatively limited. To the best of our knowledge, only a few studies have investigated this direction, most notably Transfusion~\cite{Baas2022}, FFDM~\cite{Yeh2024} and Whisfusion~\cite{Kwon2025}. On the one hand, Transfusion and FFDM formulate ASR using a multinomial diffusion process over discrete output tokens, enabling parallel decoding across time steps. On the other hand, Whisfusion combines a pretrained Whisper encoder with a pretrained text diffusion decoder, connected through a lightweight cross-attention module, with the goal of reducing the inference latency typically associated with autoregressive generation. The model is trained in a two-stage manner, where the speech encoder and text diffusion decoder are pretrained independently, followed by a separate training stage to learn their acoustic–text fusion, making the overall training pipeline more complex and less end-to-end compared to a unified ASR framework. Despite these initial efforts, diffusion-based generative NAR ASR models have been evaluated primarily on a single English-only dataset and lack comprehensive experiments on more benchmark datasets or additional ablation studies. Moreover, their recognition accuracy remains notably below that of current state-of-the-art AR ASR models, leaving their scalability and generalization capabilities insufficiently explored.

More recently, Drax~\cite{Navon2025} introduces a NAR ASR framework based on discrete flow matching (DFM)~\cite{Gat2024}, providing an alternative to diffusion-based generative modeling. In addition, Drax formulates ASR as a transport process from a uniform text distribution to the target transcript, and introduces an audio-conditioned intermediate distribution along this probability path to better align training with inference. During inference, decoding follows the learned flow trajectory through this predefined intermediate state rather than directly denoising from the source distribution. Despite the relatively competitive recognition accuracy and improved accuracy-efficiency trade-offs, its probability path design introduces additional components and hyperparameters that require careful tuning, and the intermediate distribution may still fail to capture the full variability of inference dynamics across diverse acoustic contexts.

\subsection{Iterative Masked Refinement NAR ASR}
It is worth pointing out that a few prior works have explored NAR ASR through iterative refinement of token-level transcription hypotheses, where partially observed output sequences are progressively completed and corrected across multiple decoding steps. At a high level, these models share the idea of predicting missing tokens in parallel and progressively refining the output over multiple iterations. Representative examples include Mask-CTC~\cite{Higuchi2020,Higuchi2021}, Imputer~\cite{Saharia2020}, and Align-Refine~\cite{Chi2021}. However, these methods differ from diffusion-based model in both training and inference procedures. In particular, these models typically rely on either CTC-style alignments or dynamic programming to handle variable-length outputs. Moreover, the refinement steps usually require repeated re-alignment between acoustic frames and text tokens. This reliance on explicit alignment introduces additional algorithmic complexity and constrains model design, leading to more complex training and inference process.

The most similar work to ours is~\cite{Chen2021}, which formulates ASR as an audio-conditioned masked language model trained with a BERT-style objective~\cite{Delvin2019}. A key distinction from our model lies in the underlying generative formulation. As pointed out in~\cite{Austin2021,Shi2024}, BERT-style masked reconstruction objectives, which apply a uniform loss over every noise-level without noise-dependent reweighting are not equivalent to discrete diffusion objectives and do not faithfully optimize the model likelihood. In contrast, discrete diffusion training introduces explicit reweighting $\frac{\alpha_t'}{1-\alpha_t}$ in~\eqref{eq:training_objective} to compensate for the varying expected number of masked positions across noise levels, which is derived from the forward corruption process, yielding a principled multi-step reverse-time generative model. This distinction is particularly concerning for~\cite{Chen2021}, where training follows a BERT-style objective while inference attempts to perform a multi-step iterative refinement resembling discrete diffusion. We argue that the mismatch between training and inference behaviors may lead to suboptimal solutions. In addition, their evaluation is limited to a small number of refinement steps and is conducted mainly on Japanese and Chinese datasets. As noted in~\cite{Chen2021}, performance on English and Latin alphabet based task are more challenging to their model, making it difficult to assess its robustness and general applicability across languages and decoding settings.



\section{Masked Diffusion Based ASR}
\label{sec:mdm_asr}

\subsection{Audio-conditioned Masked Diffusion Model}
Masked diffusion models can be naturally extended to ASR by conditioning the discrete denoising process on an acoustic input. In this setting, the goal is to generate a sequence of linguistic tokens that correspond to a given speech signal, while retaining the NAR advantages of discrete diffusion modeling. Let $\boldsymbol{a} \in \mathbb{R}^{T \times D}$ denote an input acoustic representation extracted by a pretrained speech encoder, where $T$ is the number of time frames and $D$ is the feature dimension. Let $\boldsymbol{x}^{(1:L)}$ denote the corresponding target token sequence. As in Section~\ref{sec:masked_diffusion}, a forward diffusion process corrupts the clean token sequence into a noisy sequence $\boldsymbol{z}_t^{(1:L)}$ at diffusion time $t$. Importantly, the forward noising process remains unchanged and is independent of the audio signal.

The distinction from conventional MDMs arises in the reverse denoising/unmasking process, which is now conditioned on the acoustic input feature $\boldsymbol{a}$. Therefore, the reverse transition is parameterized as
\begin{equation}
p_\theta(\boldsymbol{z}_s^{(1:L)} \mid \boldsymbol{z}_t^{(1:L)}, \boldsymbol{a}) = \prod_{\ell=1}^{L}
p_\theta(\boldsymbol{z}_s^{(\ell)} \mid \boldsymbol{z}_t^{(1:L)}, \boldsymbol{a}),
\label{eq:asr_reverse}
\end{equation}
where $\theta$ denotes model parameters. As before, the reverse transition is constructed by replacing the unknown clean tokens with a neural network prediction, now conditioned on both the masked-corrupted sequences and the audio signal. Specifically, let $\boldsymbol{x}^{(\ell)}_\theta(\boldsymbol{z}_t^{(1:L)}, \boldsymbol{a}) \in \Delta^{|\mathcal V|}$ denote the model-predicted categorical distribution over the vocabulary at position $\ell$, produced by an MDM decoder that attends to the acoustic representation $\boldsymbol{a}$ and the masked token sequence $\boldsymbol{z}_t^{(1:L)}$. Plugging this prediction into the posterior in~\eqref{eq:posterior} yields the audio-conditioned reverse transition
\begin{equation}
\begin{aligned}
&p_\theta(\boldsymbol{z}_s^{(\ell)} \mid \boldsymbol{z}_t^{(1:L)}, \boldsymbol{a}) \\
&=\begin{cases}
\mathrm{Cat}(\boldsymbol{z}_s^{(\ell)}; \boldsymbol{z}_t^{(\ell)}), & \boldsymbol{z}_t^{(\ell)} \neq \boldsymbol{m} \\ \mathrm{Cat}\left( \boldsymbol{z}_s^{(\ell)}; \frac{(1-\alpha_s)\boldsymbol{m} + (\alpha_s-\alpha_t)\,\boldsymbol{x}^{(\ell)}_\theta(\boldsymbol{z}_t^{(1:L)},\boldsymbol{a})}{1-\alpha_t} \right), & \boldsymbol{z}_t^{(\ell)} = \boldsymbol{m} 
\end{cases} \nonumber
\label{eq:asr_reverse_process}
\end{aligned}
\end{equation}
Following~\cite{Sahoo2024,Ou2025}, we introduce zero masking probabilities and carry-over properties for unmasked tokens to simplify the training objective in~\eqref{eq:training_objective} by excluding the timestep $t$ from the input in $\boldsymbol{x}_\theta(\cdot)$. In addition, we set the noise schedule $\alpha_t = 1-t$ as suggested in previous work \cite{Sahoo2024,Lou2024,Shi2024}. Therefore, we arrive at a modified training objective
\begin{equation}
\mathcal L = \mathbb E_q \int_{0}^{1} -\frac{1}{t} \sum_{ {\ell|\boldsymbol{z}_t^\ell=\boldsymbol{m}}} \log \left\langle \boldsymbol{x}_\theta^{(\ell)}(\boldsymbol{z}_t^{(1:L)}, \boldsymbol{a}) ,\boldsymbol{x}^{(\ell)}
\right\rangle \, \mathrm{d}t.
\label{eq:asr_objective}
\end{equation}
This objective can be interpreted as an audio-conditioned MDM language modeling loss where tokens that remain unmasked incur no loss, while masked positions are trained to directly reconstruct the original clean token sequence.

\begin{figure}
    \centering
    \includegraphics[width=\columnwidth]{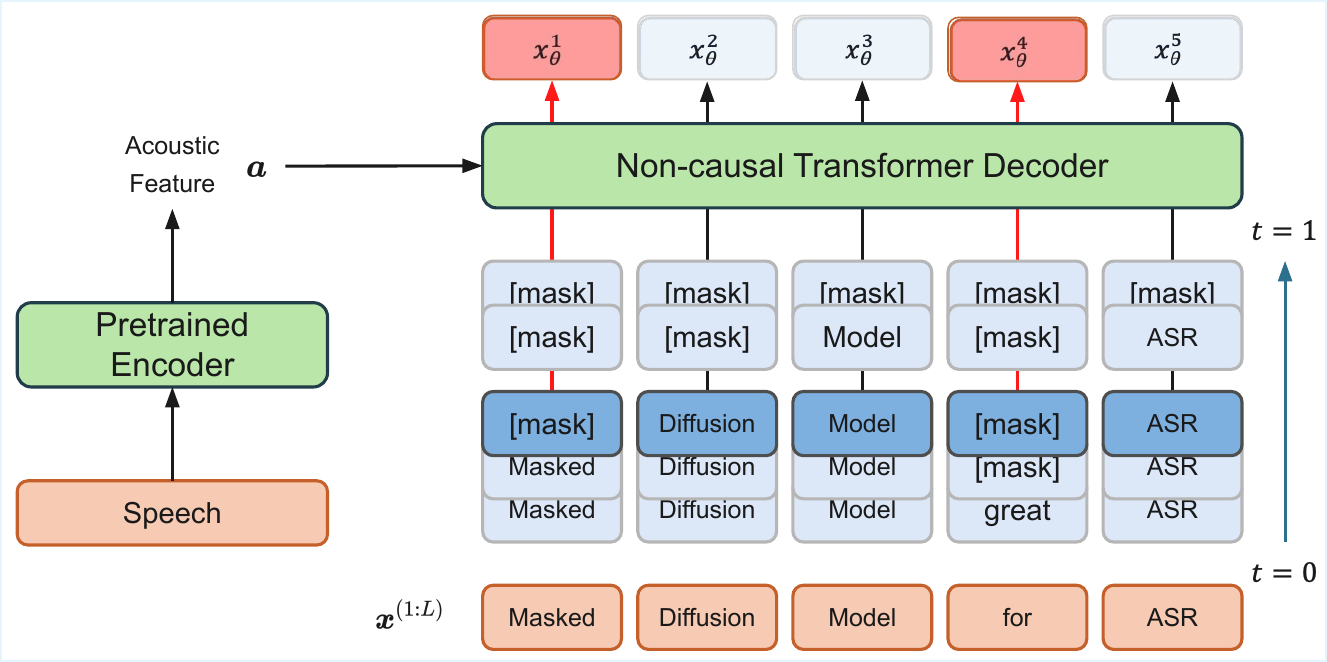}
    \caption{Overview of our MDM-ASR framework.}
    \label{fig:mdm_asr}
    \vspace{-.3cm}
\end{figure}

\subsection{Model Architecture}
Figure~\ref{fig:mdm_asr} illustrates the proposed MDM-based ASR model, referred to as MDM-ASR, which adopts a standard encoder–decoder architecture augmented with a diffusion-based decoder. The speech input is first processed by a pre-trained speech encoder to produce high-level acoustic representations, which serve as conditioning information for a Transformer-based text decoder via cross-attention. Apart from the decoding strategy, the overall architecture remains identical to conventional encoder–decoder ASR models. The key distinction lies in the decoder design. Instead of employing a causal self-attention mask as in AR decoding, the proposed decoder uses a non-causal self-attention mask, enabling each token position to attend to the entire output sequence. This bidirectional context allows the model to predict all tokens in parallel at each diffusion step, rather than generating them sequentially from left to right. As a result, the decoder can iteratively refine the full transcript at every inference step, effectively removing the sequential decoding bottleneck while preserving the representational strengths of encoder–decoder ASR models.

\subsection{Iterative Self-Correction Training}
\label{sec:isct}

A common challenge for MDMs is the mismatch between training and inference. During standard training, the model only observes oracle corrupted sequences obtained by masking the ground-truth sequence. During inference, the model instead denoises sequences that are partly self-generated. Therefore, these sequences contain plausible decoding errors from previous iterations. The discrepancy between training and inference resembles the sampling mismatch introduced by teacher forcing in AR model training~\cite{Bengio2015,Ranzato2016}, which can lead to error propagation and degraded performance.

To mitigate the mismatch between training and multi-step inference, we adopt an \textit{iterative self-correction training} (ISCT) approach inspired by~\cite{Yen2023}. ISCT explicitly simulates multiple masking/unmasking steps during training, allowing the model to learn to correct its own errors. Using two steps in the following as a proof of concept, we first sample an initial timestep $t_1$ and generate a masked sequence $\boldsymbol{z}_{t_1}$ from the ground-truth transcript $\boldsymbol{x}$. The MDM decoder produces a preliminary reconstruction $\hat{\boldsymbol{x}} = \boldsymbol{x}_\theta(\boldsymbol{z}_{t_1}, \boldsymbol{a})$, which is an estimated clean token sequence for the given speech. We then sample another timestep $t_2$ and corrupt the model’s own output by applying the forward masking process to $\hat{\boldsymbol{x}}$, yielding a partially masked sequence $\hat{\boldsymbol{z}}_{t_2} = q(\hat{\boldsymbol{z}}_{t_2}\mid \hat{\boldsymbol{x}})$ analogous to~\eqref{eq:forward_process}. Feeding $\hat{\boldsymbol{z}}_{t_2}$ into the decoder produces a refined prediction $\boldsymbol{x}_\theta(\hat{\boldsymbol{z}}_{t_2}, \boldsymbol{a})$. We can then extend the original training loss in~\eqref{eq:asr_objective} by adding a second term for this refinement step, essentially summing the cross-entropy objectives over both the first and second unmasking stages. Formally, the ISCT objective $\mathcal{L}_{\text{ISCT}}$ can be expressed as

\begin{equation}
\footnotesize
\begin{aligned}
\hspace*{-.5cm} \mathcal{L}{_\text{ISCT}} & = \mathbb E_q \int_{0}^{1} - \frac{1}{t_1} \sum_{ {\ell|\boldsymbol{z}_{t_1}^\ell=\boldsymbol{m}}} \log \left\langle \boldsymbol{x}_\theta^{(\ell)}(\boldsymbol{z}_{t_1}^{(1:L)}, \boldsymbol{a}) ,\boldsymbol{x}^{(\ell)}
\right\rangle \, \mathrm{d}t_1 \\
& + \mathbb E_q \int_{0}^{1} - \frac{1}{t_2} \sum_{ {\ell|\hat{\boldsymbol{z}}_{t_2}^\ell=\boldsymbol{m}}} \log \left\langle \boldsymbol{x}_\theta^{(\ell)}(\hat{\boldsymbol{z}}_{t_2}^{(1:L)}, \boldsymbol{a}) , \boldsymbol{x}^{(\ell)}
\right\rangle \, \mathrm{d}t_2
\label{eq:isct_objective}
\end{aligned}
\end{equation}

It is important to note that prior flow-matching models~\cite{Shaul2025,Lim2025,Cross2025}, including Drax~\cite{Navon2025}, addresses the training–inference mismatch by introducing a fixed audio-conditioned intermediate distribution along a predefined probability path. However, this method tightly couples the training dynamics with the chosen inference path, making design choices difficult and non-trivial. In contrast, ISCT directly exposes the model to its own intermediate decoding states. By corrupting and refining the model’s self-generated predictions during training, the MDM decoder learns to correct realistic errors that arise during iterative inference, rather than relying on a handcrafted or static intermediate representation. This data-driven process allows the model to capture a broader range of inference dynamics induced by acoustic variability, early mispredictions, and domain mismatch. As a result, ISCT provides a closer alignment between training and inference in the proposed MDM-ASR, leading to improved robustness and generalization across diverse speech conditions.

\subsection{Inference Samplers}
\label{subsec:samplers}

As discussed in Section~\ref{sec:intro}, a key advantage of MDMs over AR models lies in their flexibility during inference. Unlike AR decoding, which follows a fixed left-to-right order, MDMs allow tokens to be decoded in arbitrary orders. Moreover, multiple tokens can be decoded in parallel to trade off accuracy for efficiency. This flexibility enables a variety of decoding strategies beyond the vanilla random-unmasking sampler~\cite{Sahoo2024}, motivating a series of works that develop more advanced sampling methods~\cite{Kim2025,yu2025dimple,Ben2025,huang2025pc}. In this section, we describe the inference samplers evaluated in our experiments, including our proposed combination of entropy-bounded decoding and positional bias.

\subsubsection{Discrete Flow-Matching Sampler}

Although our ASR system is based on MDMs, we can still apply the Discrete Flow-Matching (DFM) sampler at inference time by interpreting the model’s categorical predictions as a probability flow in the discrete token space. Although applying DFM in this setting is not entirely straightforward, for completeness and to ensure a fair comparison with flow-matching based baselines, we also evaluate this sampler in our experiments. At each decoding step, the predicted distribution $\boldsymbol{x}^{(1:L)}_\theta(\boldsymbol{z}_t, \boldsymbol{a})$ over clean tokens is used to construct a discrete flow field that guides the evolution of tokens toward the predicted clean state. Let $\boldsymbol{z}_t^{(\ell)} \in \{0, 1\}^{\mathcal{|V|}}$ denote as a one-hot vector representing the current token at time $t$ at position $\ell$. The flow direction is defined as
\[
\boldsymbol{v}^{(\ell)} = \frac{\boldsymbol{x}^{(\ell)}_\theta(\boldsymbol{z}_t, \boldsymbol{a}) - \boldsymbol{z}_t^{(\ell)}}{1 - \alpha_t}
\]
which represents a normalized drift from the current discrete state toward the predicted clean-token distribution at position $\ell$. For a target time $s$ with $0 < s < t$, the categorical distribution at time $s$ is obtained via a linear update
\[
p_\theta(\boldsymbol{z}_s^{(\ell)} \mid \boldsymbol{z}_t^{(1:L)}, \boldsymbol{a}) = \boldsymbol{z}_t^{(\ell)} + (\alpha_s - \alpha_t)\,\boldsymbol{v}^{(\ell)}
\]
from which $\boldsymbol{z}_s^{(\ell)}$ are sampled to form the next decoding state.

\subsubsection{Confidence-based Samplers}

While random unmasking mirrors the training-time masking procedure, prior studies have shown that the inference-time unmasking order can have a substantial impact on sampling quality, especially when prediction errors are present~\cite{Kim2025}. In such cases, selecting positions to unmask based on model-derived criteria, such as confidence, entropy, or margin computed from the predicted token distributions, consistently outperforms vanilla random unmasking~\cite{Kim2025,Ben2025,Ku2026}.

One common class of confidence-based samplers is the Confidence Top-$K$ (Conf-Top-$K$) sampler~\cite{Kim2025}. At each decoding iteration, the model first computes token distributions for all currently masked positions and ranks them according to a confidence criterion. Token positions are then ranked by the maximum predicted probability, and the top $K$ position are selected for unmasking. The parameter $K$ controls an efficiency-accuracy tradeoff: larger values of $K$ reduce the number of function evaluations (NFEs) but may degrade decoding quality.

To improve efficiency without sacrificing quality, the Entropy-Bounded Confidence sampler (EB-Conf)~\cite{Ben2025} replaces the fixed $K$ with an adaptive selection rule. After sorting masked positions by confidence, EB-Conf selects the largest prefix set $U$ whose predicted entropies satisfy an entropy-budget constraint,
\begin{equation}
\sum_{l \in U} H\left(\boldsymbol{x}^{(\ell)}_\theta(\boldsymbol{z}_t^{(1:L)}, \boldsymbol{a})\right)-\max _{l \in U} H\left(\boldsymbol{x}^{(\ell)}_\theta(\boldsymbol{z}_t^{(1:L)}, \boldsymbol{a})\right) \leq \gamma
\end{equation}
and then unmasks all positions in $U$ in the current decoding step. Intuitively, when the model is simultaneously confident about many positions (low entropy), EB-Conf unmasks more tokens per iteration to reduce NFEs. However, when there is higher uncertainty, it automatically unmasks fewer tokens, mitigating the quality degradation observed with large fixed-$K$ parallel unmasking. The threshold $\gamma$ directly controls this behavior, with $\gamma=0$ reducing to single-token unmasking and $\gamma\rightarrow\infty$ allowing the unmasking of all remaining tokens in one decoding step. 

\subsubsection{Position-Biased EB-Conf Sampler}

In addition to entropy-bounded selection, we propose to incorporate the positional trajectory prior proposed in~\cite{huang2025pc} to further improve decoding performance. Specifically, we introduce a positional bias term $P_i = e^{-\lambda i}$, where $i$ denotes the token index and $\lambda$ controls the strength of positional regularization. After computing the confidence scores $c_i$ for each masked position, we reweight them using the positional bias, yielding adjusted scores $\hat{c}_i = P_i \cdot c_i$. This mechanism encourages earlier positions to be decoded first. The entropy-bounded criterion is then applied to the re-ranked positions to determine the adaptive unmasking set. We refer to this combined strategy as the \emph{PBEB-Conf} sampler.

While EB-Conf~\cite{Ben2025} and the positional trajectory prior~\cite{huang2025pc} were originally proposed as separate techniques, to the best of our knowledge, this is the first work to integrate entropy-bounded adaptive decoding with positional bias. Empirically, this combination consistently improves decoding performance, as demonstrated in Section~\ref{subsubsec:sampler_choice}.

\section{Experimental Setup}
\label{sec:exp_setup}
\subsection{Datasets \& Evaluation Metrics}
\subsubsection{Datasets}
For English-only experiments, we evaluate our models on four datasets from the Hugging Face Open ASR benchmark: LibriSpeech~\cite{Panayotov2015}, Earnings22~\cite{Delrio2022}, AMI~\cite{Carletta2007,Renals2007}, and VoxPopuli~\cite{Wang2021}. LibriSpeech is a widely used large-scale benchmark comprising approximately 1,000 hours of read English speech derived from audiobooks, with standard evaluation splits including \texttt{test-clean} and \texttt{test-other}. Earnings22 consists of 119 hours of earnings call recordings featuring speakers from a diverse set of global companies. The dataset is designed to capture substantial variability in speakers, accents, and financial discourse, thereby reflecting realistic domain conditions. The AMI Meeting Corpus contains approximately 100 hours of multi-party meeting recordings collected across multiple recording channels. It provides manually produced orthographic transcriptions aligned at the word level, making it well suited for detailed ASR evaluation in conversational settings. The VoxPopuli dataset is a large-scale multilingual speech corpus derived from European Parliament recordings, providing audio in over 20 languages. Following~\cite{Gandhi2022}, we filter and partition each dataset into training, validation, and test sets, as summarized in Table~\ref{tab:en_data}.
For multilingual studies, we use the commonly adopted Multilingual LibriSpeech (MLS)~\cite{Pratap2020} dataset. We consider three non-English languages that are covered by the pretrained model, namely German (DE), Spanish (ES), and French (FR). The total amount of training data is approximately 4,000 hours. Detailed dataset statistics are summarized in Table~\ref{tab:mls_data}.

\begin{table}[t!]
    \centering
    \caption{English datasets description and statistics, including train, validation, and test splits in hours of audio data.}
    \label{tab:en_data}
    \adjustbox{width=\columnwidth}{
    \begin{tabular}{l|c|c|c|c}
        \toprule
        \textbf{Dataset} & \textbf{Domain} & \textbf{Train (h)} & \textbf{Val (h)} & \textbf{Test (h)} \\ \midrule
        LibriSpeech & Audiobook & 960 & 11 & 10.74 \\
        Earnings22 & Financial meetings & 105 & 5 & 5.43 \\
        AMI & Meetings & 78 & 5 & 8.54 \\
        VoxPopuli & European parliament & 523 & 5 & 4.93 \\
        \bottomrule
   \end{tabular}
   }
\end{table}

\begin{table}[t!]
    \centering
    \caption{MLS dataset description and statistics, including the 3 considered languages: German, Spanish, and French.}
    \label{tab:mls_data}
    \adjustbox{width=0.7\columnwidth}{
    \begin{tabular}{l|c|c|c}
        \toprule
        \textbf{Language} & \textbf{Train (h)} & \textbf{Val (h)} & \textbf{Test (h)} \\ \midrule
        German & 1966.51 & 14.28 & 14.29 \\
        Spanish & 917.68 & 9.99 & 10 \\
        French & 1076.58 & 10.07 & 10.07 \\
        \bottomrule
   \end{tabular}
   }
\end{table}

\subsubsection{Evaluation Metrics}

We evaluate our models in terms of both recognition accuracy and decoding efficiency. Recognition accuracy is measured using word error rate (WER). Prior to WER computation, both ground-truth transcripts and model predictions are normalized using the Whisper Normalizer~\cite{Radford2023}. Decoding efficiency is quantified by the inverse real-time factor (RTFx), defined as the ratio between the total duration of the audio and the total decoding time
\begin{equation}
\text{RTFx} = \frac{\text{Total audio duration (s)}}{\text{Decoding time (s)}} .
\end{equation}
An RTFx value greater than 1 indicates faster-than-real-time decoding, with larger values corresponding to higher efficiency. For a fair comparison, all models are evaluated on a single NVIDIA A100 GPU with a batch size of 1, using full-precision inference and without any model compilation or runtime optimization. Runtime results on the LibriSpeech \texttt{test-clean} and \texttt{test-other} sets are averaged over all utterances.


\begin{table*}[t]
\centering
\caption{Evaluation results on four English benchmark test sets: LibriSpeech test-clean (LS Clean), LibriSpeech test-other (LS Other), Earnings22, AMI, and VoxPopuli. We report word error rate (WER), and decoding efficiency measured by RTFx. WERs for most baseline models are reproduced from the HF leaderboard, while models marked with $^\dag$ are obtained directly from official reports or peer-reviewed literature. RTFx for all models are computed using the corresponding public-released checkpoints. Avg. refers to the average WER of the five test sets.}
\label{tab:en_results}
\begin{tabular}{clcccccccc}
\toprule
& \multirow{2}{*}{\textbf{Model}} & \multicolumn{6}{c}{\textbf{WER} (\%)$\downarrow$} & \multirow{2}{*}{\textbf{Params (B)}} & \multirow{2}{*}{\textbf{RTFx}$\uparrow$} \\
& & LS Clean & LS Other & Earnings22 & AMI & VoxPopuli & Avg. \\
\midrule
\multirow{6}{*}{\rotatebox[origin=c]{90}{\textbf{AR}}}
& Whisper-large-v3 & 2.0 & 3.9 & 11.3 & 16.0 & 9.5 & 8.5 & 1.5 & 12.83 \\
& OWSM-v3.1$^\dag$ & 2.4 & 5.0 & 15.4 & 20.4 & 8.4 & 10.3 & 1.0 & 15.52 \\
& Canary-1b-flash & 1.5 & 2.9 & 12.8 & 13.1 & 5.6 & 7.2 & 1.0 & 29.25 \\
& Phi-4-multimodal & 1.7 & 3.8 & 10.2 & 11.7 & 6.0 & 6.7 & 5.6 & 4.85 \\
& Qwen2-Audio$^\dag$ & 1.6 & 3.6 & 14.1 & 15.2 & 7.1 & 8.3 & 8.4 & 4.18 \\
& Voxtral-Mini & 1.9 & 4.1 & 10.7 & 16.3 & 7.1 & 8.0 & 3.0 & 12.19 \\
\midrule
\multirow{7}{*}{\rotatebox[origin=c]{90}{\textbf{NAR}}}
& OWSM-CTC$^\dag$ & 2.4 & 5.2 & 17.2 & 23.8 & 8.6 & 11.4 & 1.0 & 132.54 \\
& Parakeet-CTC & 1.8 & 3.5 & 13.8 & 15.7 & 6.6 & 8.3 & 1.1 & 120.20 \\ 
\cmidrule{2-10}
& TransFusion$^\dag$ & 6.7 & 8.8 & - & - & - & - & 0.2 & 0.57 \\
& Whisfusion$^\dag$ & 8.3 & 17.0 & - & - & - & - & 0.3 & 84.86 \\
& FDDM$^\dag$ & 4.0 & 7.2 & - & - & - & - & 0.6 & - \\
& Drax$^\dag$ & 2.6 & 5.7 & 15.2 & 13.9 & 8.6 & 9.2 & 1.2 & 11.32 \\ \cmidrule{2-10} \\ [-1.4em] \cmidrule{2-10}
& MDM-ASR (ours) & 1.8 & 3.6 & 10.7 & 12.2 & 6.0 & 6.9 & 1.0 & 46.81 \\
\bottomrule
\end{tabular}
\end{table*}

\subsection{Training \& Inference Setup}
All experiments are conducted using Canary encoder-decoder ASR models implemented in NVIDIA NeMo~\cite{Zelasko2025,Kuchaiev2019,Harper2023}. Our MDM-based ASR system is initialized from the official pretrained \texttt{canary-1b-flash} checkpoint, which contains approximately 1B trainable parameters. Since \texttt{canary-1b-flash} is originally trained as an AR model, we modify the decoder self-attention mechanism by replacing causal attention with non-causal attention to support masked diffusion decoding. Detailed information about Canary architecture can be found in~\cite{Zelasko2025,Rekesh2023}.

Training is performed using the Adam optimizer~\cite{Kingma2015} with a learning rate of $5\times10^{-4}$ and a batch size of 8. All experiments adopt a two-step ISCT schedule, although the proposed approach naturally extends to a larger number of steps.
During training, shorter sequences are padded with end-of-sentence (\texttt{EOS}) tokens to achieve uniform sequence lengths, treating \texttt{EOS} as a standard token. During inference, the model may generate multiple \texttt{EOS} tokens near the end of the sequence. Therefore, we truncate the output at the first \texttt{EOS} token and discard all subsequent positions. This strategy prevents redundant refinement of positions beyond the underlying audio content and improves decoding stability and efficiency without requiring explicit length prediction or additional heuristics~\cite{Ghazvininejad2019}. Decoding is initialized with a maximum sequence length of 256 tokens, which is sufficient for standard ASR benchmarks.

\subsection{Baselines}
\label{sec:baseline}
We compare our model against a diverse set of top-tier ASR models from the HuggingFace (HF) Open ASR Leaderboard~\cite{Srivastav2025}, spanning large-scale AR, NAR, and diffusion-based models. Whisper~\cite{Radford2023} is a multilingual encoder–decoder model trained on approximately 5 million hours of weakly supervised speech–text data. Canary-1b-flash~\cite{Zelasko2025} is a family of multi-tasking models based on Canary architecture~\cite{Puvvada2024} that supports four languages (English, German, French, Spanish). SeamlessM4T-large-v2~\cite{Communication2023} is a large-scale multilingual and multimodal sequence-to-sequence model that supports speech recognition across many languages, trained end-to-end on massive multilingual speech and text data. Phi-4-multimodal~\cite{Microsoft2025} is a lightweight open multimodal foundation model that leverages the language, vision, and speech research and datasets used for Phi-3.5 and 4.0 models. Qwen2-Audio~\cite{Chu2024} extends the Qwen large language model~\cite{Chu2023} with an audio encoder to enable speech understanding and recognition. Voxtral~\cite{Liu2025} is built on the Mistral LLM~\cite{Jiang2023} backbone and integrates a dedicated speech encoder for ASR.

For NAR models, we include strong CTC-based baselines. OWSM-CTC~\cite{Peng2024} is a fully open speech model trained on up to three million hours of curated English speech. Parakeet-CTC uses an XXL version of FastConformer~\cite{Rekesh2023}, and is currently the best CTC-based model on the HF leaderboard. XLSR-53~\cite{Conneau2020,Babu2022} is a multilingual speech representation model based on Wav2Vec2.0, pretrained in a self-supervised manner on raw audio from 53 languages and commonly fine-tuned with a CTC objective for multilingual ASR. We also include the diffusion- and flow-matching based baselines as mentioned in Section~\ref{sec:nar_asr}, namely, Transfusion~\cite{Baas2022}, Whisfusion~\cite{Kwon2025}, FFDM~\cite{Yeh2024}, and Drax~\cite{Navon2025}. Together, these baselines span a broad spectrum of modern ASR models, including large encoder–decoder models, LLM-augmented models, as well as CTC- and diffusion-based NAR methods. We note that direct comparisons under identical training conditions are often infeasible due to the nature of large-scale foundation models. For instance, Whisper-large-v3 is trained on millions of hours of audio using proprietary data, undisclosed training pipelines, and substantial computational resources, making it impractical for us to reproduce or retrain under controlled and comparable settings. Therefore, following~\cite{Navon2025} and prior literature, we evaluate our model against representative state-of-the-art ASR models to the best of our knowledge, which are also among the top-ranked models on the HF leaderboard. We also observe discrepancies among the results reported across prior literature. To address this issue and ensure a consistent and reproducible evaluation results, we adopt the official scripts from~\cite{Srivastav2025} to reproduce the baseline results, thereby aligning the evaluation setup for a fair comparison.

\section{Results}
\label{sec:exp_results}

\subsection{English Benchmark Results}
\label{sec:en_results}
Table \ref{tab:en_results} reports the results on the English-only benchmarks. Unless otherwise specified, MDM-ASR results are reported for our best-performing 1B model configuration, trained with ISCT, and decoded using the proposed PBEB-Conf sampler with $\lambda=0.2, \gamma=0.05$ and a maximum NFE of 32. Among generative NAR models, the proposed MDM-ASR establishes a new state-of-the-art to the best of our knowledge. Earlier diffusion-based ASR models, including TransFusion, Whisfusion, and FFDM, either lack decoding efficiency, e.g., RTFx of 0.57 for Transfusion, or exhibit limited recognition accuracy, e.g., WER of 8.3\% on LS Clean and 17\% on LS Other for Whisfusion. Even compared with Drax, the strongest prior generative NAR baseline, MDM-ASR still achieves consistently superior performance across all benchmarks with faster decoding speech, around \textbf{4.1x} speedup in RTFx at a similar model size. On LibriSpeech, our model achieves 1.8\% WER on LS Clean and 3.6\% on LS Other, improving upon Drax which reports 2.6\% and 5.7\% WER respectively, corresponding to a relative WER improvement of approximately 31\% on LS Clean and 37\% on LS Other. Similar trends are observed on Earnings22, AMI, and VoxPopuli. MDM-ASR reduces WER from 15.2\% to 10.7\% for Earnings22, from 13.9\% to 12.2\% for AMI, and on VoxPopuli, from 8.6\% to 6.0\%, further indicating improved generalization to conversational and real-world speech. These results demonstrate that the proposed model yields a more effective and efficient alternative to existing generative NAR ASR frameworks.

\begin{table}[t!]
\centering
\caption{Multilingual evaluation results on the MLS dataset, including German (DE), Spanish (ES), and French (FR). Results are reproduced from the HF leaderboard, while models marked with $^\dag$ are obtained directly from official reports or peer-reviewed literature.}
\label{tab:mls_results}
\begin{tabular}{clccc}
\toprule
& \multirow{2}{*}{\textbf{Method}} & \multicolumn{3}{c}{\textbf{WER} (\%)$\downarrow$} \\
& & DE & ES & FR \\
\midrule
\multirow{6}{*}{\rotatebox[origin=c]{90}{\textbf{AR}}}
& Whisper-larve-v3 & 3.1 & 3.0 & 4.8 \\
& OWSM-v3.1$^\dag$ & 10.8 & 9.0 & 12.1 \\
& SeamlessM4T-large-v2$^\dag$ & 6.1 & 4.1 & 5.4 \\
& Canary-1b-flash & 4.4 & 2.7 & 4.5  \\
& Phi-4-multimodal & 4.8 & 3.6 & 4.9 \\
& Voxtral-Mini & 7.1 & 5.1 & 5.3 \\
\midrule
\multirow{5}{*}{\rotatebox[origin=c]{90}{\textbf{NAR}}}
& OWSM-CTC$^\dag$ & 11.9 & 10.3 & 12.9 \\
& XLSR-53$^\dag$ & 6.5 & 6.1 & 5.6 \\
\cmidrule{2-5}
& Drax$^\dag$ & 7.7 & 5.4 & 7.1 \\ \cmidrule{2-5} \\ [-1.4em] \cmidrule{2-5}
& MDM-ASR (ours) & 3.6 & 2.9 & 3.8 \\
\bottomrule
\end{tabular}
\vspace{-.2cm} 
\end{table}

When compared with AR methods, we observe that although strong AR models remain competitive in absolute recognition accuracy, the gap between AR and NAR ASR is markedly reduced relative to prior NAR models. In particular, MDM-ASR achieves comparable or even better performance on several challenging benchmarks, surpassing Whisper-large-v3 on LS Other (3.6\% to 3.9\%), Earnings22 (10.7\% to 11.3\%), and AMI (12.2\% to 16.0\%), and outperforming OWSM-v3.1 on all five English test sets. Even compared with the strongest AR model Canary-1b-flash, MDM-ASR achieves better average WER of 6.9\% over 7.2\%. In addition, MDM-ASR achieves substantially higher decoding efficiency than all AR baselines, delivering approximately \textbf{3.6x} speedup over Whisper and \textbf{3.0x} over OWSM. Even when compared with the fastest AR model, Canary-1b-flash, MDM-ASR still provides an efficiency gain of about \textbf{1.6x}. Notably, Whisper and OWSM are trained on several million hours of weakly supervised speech–text data, whereas MDM-ASR is trained on substantially less curated speech data, making these results particularly promising. A similar trend is observed when comparing against recent LLM-augmented ASR models, including Phi-4-multimodal, Qwen2-Audio, and Voxtral-Mini. These models benefit from strong pretrained large language model priors and extensive textual knowledge, which naturally enhance recognition performance. Despite this, MDM-ASR demonstrates competitive accuracy across diverse English benchmarks, further underscoring its effectiveness in closing the performance gap between NAR and AR ASR models.

For completeness, we also compare MDM-ASR with strong CTC-based ASR models, which represent the most widely adopted NAR paradigm in practice. CTC-based models such as OWSM-CTC and Parakeet-CTC offer extremely fast decoding (RTFx $> 100$) due to their single-pass inference. Among them, Parakeet-CTC achieves competitive performance on LibriSpeech, reflecting the strength of large-scale CTC models on clean read speech. However, on benchmarks such as Earnings22, AMI, and VoxPopuli, MDM-ASR consistently outperforms Parakeet-CTC. In general, MDM-ASR provides a more balanced accuracy profile across diverse benchmarks, achieving lower WER while retaining the parallel decoding advantages of NAR inference.

\begin{figure}[t!]
    \centering
    \includegraphics[width=0.8\columnwidth]{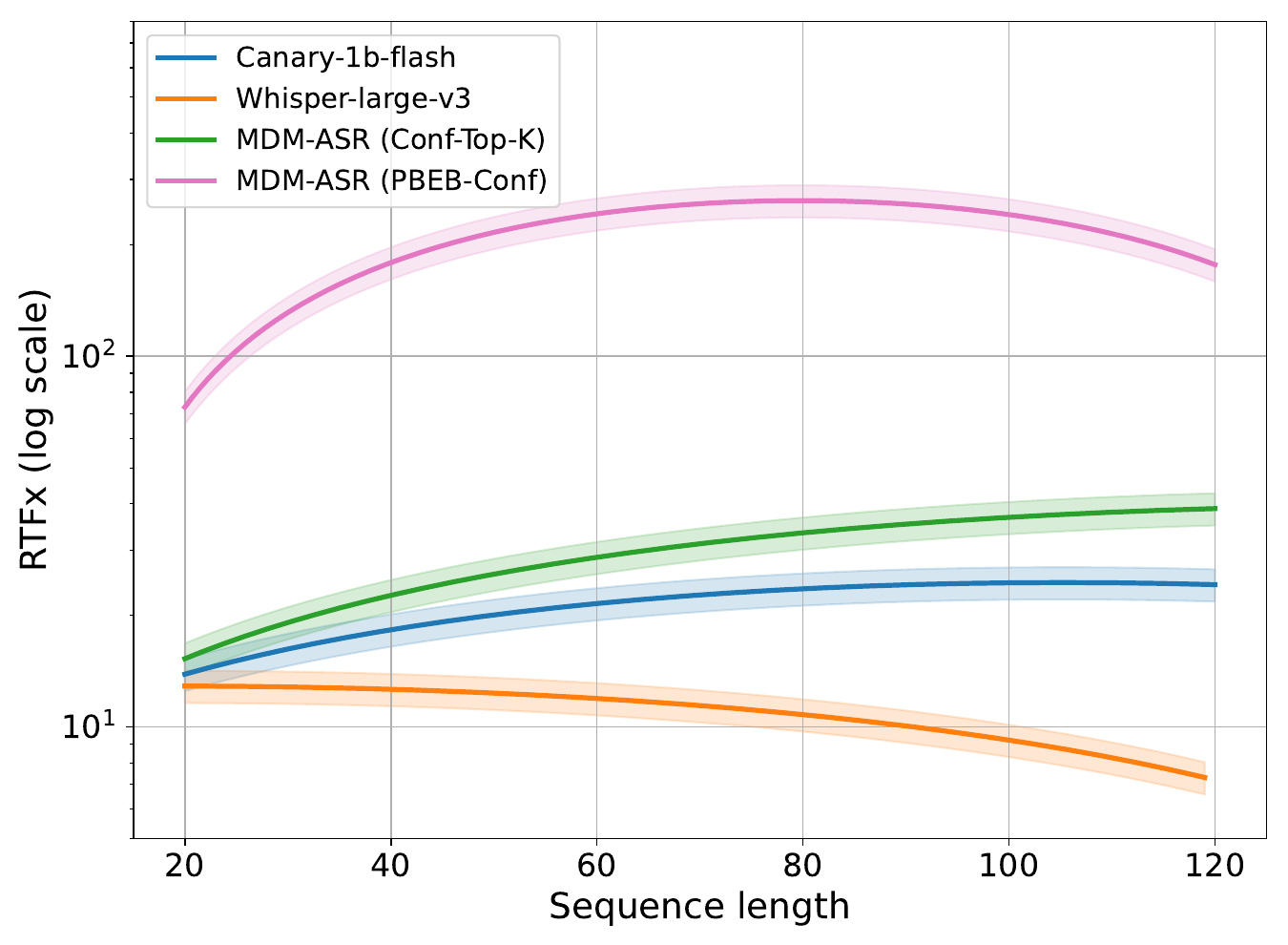}
    \caption{The RTFx as a function of sequence length. Shaded regions indicate a $\pm10\%$ variability band around the fitted curve to illustrate the uncertainty of the estimated trend.}
    \label{fig:aet}
\end{figure}

\subsection{Multilingual Results}
\label{sec:ml_results}
Table \ref{tab:mls_results} reports multilingual recognition results on MLS for German, Spanish, and French. Overall, MDM-ASR works effectively with multilingual scenario and achieves competitive accuracy against strong AR and NAR baselines. In particular, MDM-ASR attains 3.6\%/2.9\%/3.8\% WER on German/Spanish/French, substantially improving over the strongest prior generative NAR baseline Drax (7.7\%/5.4\%/7.1\%), the CTC-based models, such as OWSM-CTC (11.9\%/10.3\%/12.9\%), and XLSR-53 (6.5\%/6.1\%/5.3\%). When compared with AR models, MDM-ASR is on-par or exceeds several widely used large-scale models. For instance, it outperforms Whisper-large-v3 on Spanish (2.9\% vs. 3.0\%) and French (3.8\% vs.4.8\%), and significantly surpasses SeamlessM4T-large-v2 and OWSM-v3.1 across all three languages. Notably, MDM-ASR also achieves lower WER than recent LLM-augmented ASR models, including Phi-4-multimodal (4.8\%/3.6\%/4.9\%) and Voxtral-Mini (7.1\%/5.1\%/5.3\%), across all three languages. We note that these large-scale multilingual models, such as Whisper, OWSM, Phi-4-multimodal, and Voxtral, are trained to support a much broader set of languages, resulting in substantially larger vocabularies and more diverse modeling objectives, which may partially affect performance when evaluated on a small subset of languages. Nevertheless, when compared under a more closely matched setup, namely Canary-1b-flash, which shares an identical encoder architecture and training paradigm, we still observe meaningful empirical gains on German (3.6\% vs. 4.4\%) and French (3.8\% vs. 4.5\%). These results suggest that the improvements achieved by MDM-ASR are not solely attributable to differences in language coverage, but rather reflect the effectiveness of masked diffusion decoding for multilingual ASR within a NAR framework.

\subsection{Decoding Efficiency}
\label{subsec:decoding_efficiency}
The key advantage of our proposed MDM-ASR over AR models lies in its decoding efficiency. Figure~\ref{fig:aet} shows how decoding speed, measured by RTFx, scales with output sequence length. As sequence length increases, AR models such as \mbox{Whisper-large-v3} exhibit steadily declining efficiency due to their inherently sequential, token-by-token decoding process, while Canary-1b-flash provides only limited scaling benefits but remains largely constant. In contrast, MDM-ASR maintains substantially higher RTFx across all sequence lengths, with the performance gap widening for longer outputs. All confidence-based decoding variants further improve efficiency, and the entropy-bounded strategies, including EB-Conf and PBEB-Conf, consistently achieves the highest RTFx, outperforming AR baselines by a large margin throughout the entire range. For simplicity we present results on PBEB-Conf, while noting both methods demonstrate comparable speed. The slight decline in PBEB-Conf at longer sequence lengths stems from its adaptive update strategy, causing it to converge toward Conf-Top-$K$ performance, which is still faster than AR decoding. Overall, these results demonstrate the favorable scaling properties of masked diffusion decoding, which avoids the linear latency growth inherent to AR models and enables consistently faster-than-real-time performance, especially for long-form utterances where decoding efficiency is most critical.



\subsection{Ablation Studies}

\begin{figure}[t!]
    \centering
    \includegraphics[width=\columnwidth]{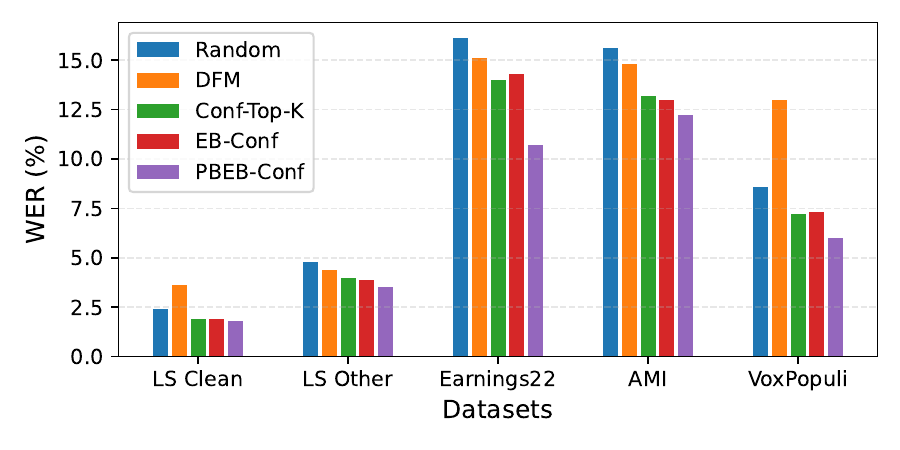}
    \caption{WER comparison between different samplers.}
    \label{fig:sampler_wer_comparison}
    \vspace{-.3cm} 
\end{figure}

\subsubsection{Effect of Sampler Choice}
\label{subsubsec:sampler_choice}

In Section~\ref{subsec:samplers}, we introduced five inference samplers: Random-Unmasking, DFM, Conf-Top-$K$, EB-Conf, and PBEB-Conf. We compare their decoding performance on five English test sets, as summarized in Figure~\ref{fig:sampler_wer_comparison}. Overall, the Random-Unmasking and the DFM sampler yields clearly suboptimal performance across all test sets compared to the other three methods, especially on the more challenging Earnings22 and AMI datasets. This behavior may partially explain the superior performance of our model compared to Drax, since the DFM sampler also selects unmasking positions in a largely random manner. In contrast, the confidence-based samplers achieve similar and consistently strong performance. This observation aligns with prior work~\cite{Kim2025,Ben2025,yu2025dimple,Ku2026} and indicates that the model’s confidence estimates serve as a reliable proxy for decoding accuracy. Furthermore, adopting the EB-Conf sampler maintains decoding accuracy while improving inference speed, as discussed in Section~\ref{subsec:decoding_efficiency}. Finally, incorporating the positional bias term yields additional performance gains: the PBEB-Conf sampler consistently improves results.

\begin{figure}
    \centering
    \includegraphics[width=0.9\columnwidth]{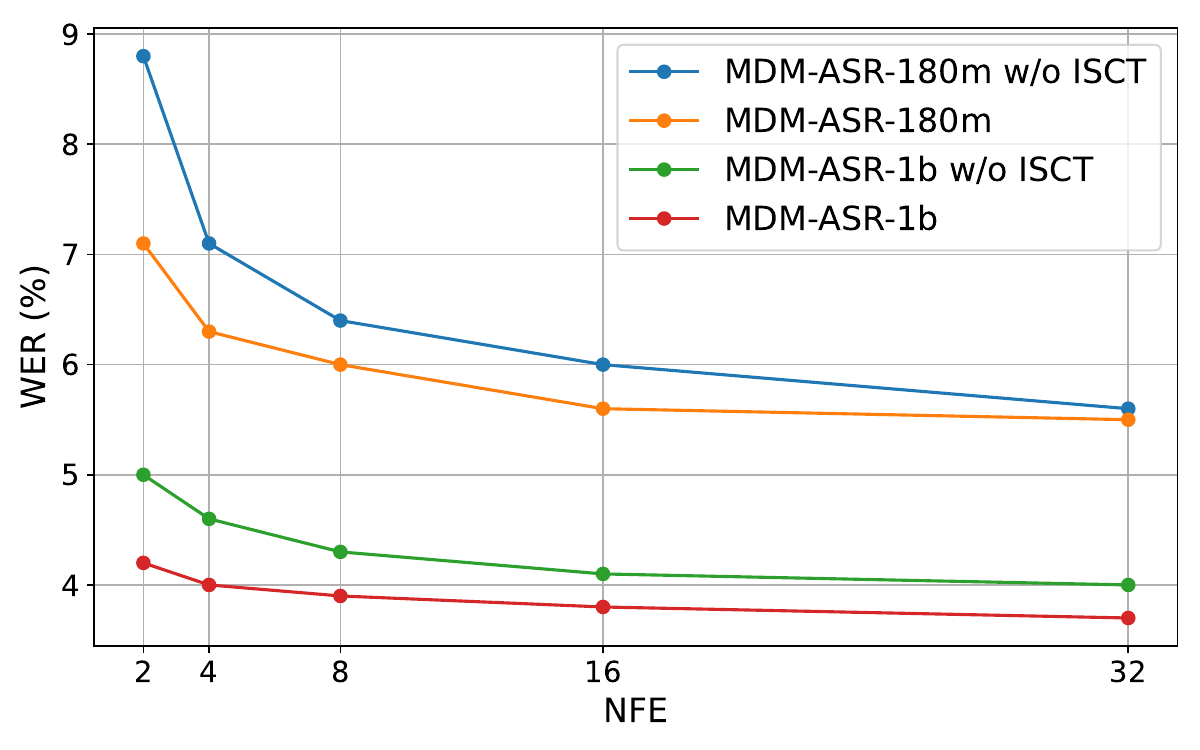}
    \caption{Effect of the proposed ISCT. WER comparison on LS Other for 180M and 1B models under varying maximum NFEs.}
    \label{fig:unf}
    \vspace{-.3cm} 
\end{figure}

\subsubsection{Effect of Iterative Self-Correction Training}
Figure~\ref{fig:unf} shows the effect of ISCT on WER across different NFEs for the MDM-ASR model with two different model sizes: MDM-ASR-180m with 180M parameters and MDM-ASR-1b with 1B parameters. For simplicity, we present results on LS Other, while noting that a similar trend is observed on LS Clean. For both model sizes, ISCT consistently improves recognition accuracy across all decoding budgets, with the largest gains observed at smaller numbers of decoding steps (2, 4, and 8), where early prediction errors are more difficult to correct through iterative refinement alone. This indicates that exposing the model to its own intermediate predictions during training effectively alleviates the mismatch between training and inference in masked diffusion decoding. While the relative improvements are smaller for the 1B model, they remain consistent across decoding steps and are particularly evident on the more challenging LS Other set. Notably, these gains are achieved without additional inference cost, highlighting ISCT as a practical and effective strategy for improving both robustness and generalization in masked diffusion–based ASR.


\section{Discussions and Conclusion}
\textbf{Limitations:} While the proposed MDM-ASR framework demonstrates competitive accuracy and favorable decoding efficiency, several limitations remain. Our evaluation is limited to a subset of publicly available datasets, and extending experiments to more diverse languages, domains, and real-world conditions, remains an important direction for future work. In addition, our results are based on only a specific subset of possible design choices. Investigating alternative encoders, adaptive masking schedules, advanced training techniques, and multiple steps of ISCT configurations may further improve performance and robustness but is left for future exploration. Finally, while our work focuses on establishing the general performance of MDM-ASR on standard ASR benchmarks, we note that this framework can be extended to other ASR applications, including streaming ASR and domain adaptation, which we consider promising directions for future work.

\noindent \textbf{Conclusion:} We have presented MDM-ASR, a simple audio-conditioned masked diffusion framework for non-autoregressive ASR that preserves the standard encoder-decoder architecture while replacing sequential left-to-right decoding with iterative parallel unmasking. To better align training and inference, we proposed ISCT and investigated practical inference samplers that improve stability and efficiency. Experiments on English and multilingual benchmarks show that MDM-ASR consistently outperforms prior generative NAR models and substantially narrows the gap to strong autoregressive models, while providing favorable efficiency in decoding speed. We further conduct comprehensive ablation studies. Together, these results establish masked diffusion as a competitive and practical direction for effective and efficient NAR ASR. We will open our source code and all hyperparameters with pre-trained models in the future.


\clearpage
\bibliographystyle{IEEEtran}
\bibliography{mybib}

\end{document}